# Magnetically tunable terahertz all-dielectric metamaterial

Chuwen Lan, Hailian Wu, Daquan Yang, Zehua Gao*

Beijing Laboratory of Advanced Information Networks & Beijing Key Laboratory of Network System Architecture and Convergence Acknowledgment, School of Information and Communication Engineering, Beijing University of Posts and Telecommunications, Beijing 100876, China

*gaozehua@bupt.edu.cn*

*Abstract:*

All-dielectric metamaterials composed of high index and low loss dielectric resonators have become a promising way for high-efficient optical devices. However, fabricating terahertz all-dielectric metamaterials and actively tuning their properties still remain challenges. In this paper, an effective method has been developed to prepare high-quality microspheres based all-dielectric metamaterial operating in terahertz (THz) range. Then, we propose a magnetically tunable THz all-dielectric metamaterial based on liquid crystal (LC). The all-dielectric metamaterial is immersed into LC and tuned by external magnetic field. We show that the induced Mie resonances can be effectively tuned by external magnetic field and the tunability is highly sensitive to the directors of applied magnetic field. A good tunability performance is obtained for the first electric resonance. This work provides a new method for high-quality THz all-dielectric metamaterials and paves a promising way for tunable THz all-dielectric metamaterial, which would find considerable applications in THz devices.

*Main Text:*

The term "metamaterials" or "metasurfaces" refers to artificially structured materials composed of arrayed subwavelength structures, which are engineered to obtain unusual electromagnetic properties in a desired frequency range. Unlike conventional materials, their electromagnetic properties are mainly determined by the geometrical parameters and packing arrangement of unit cells, rather than the material composition. Based on this feature, one can tailor their electromagnetic properties at the subwavelength level. Consequently, various exotic properties like negative refractive index [1], near- zero refractive index [2] and extremely large refractive index [3], as well as considerable applications including invisibility cloak [4], superlens [5] and perfect absorber [6] are demonstrated at different spectral regions. Traditionally, subwavelength metallic structures like split ring resonators (SRRs), rods and disks are used as the building blocks of the metamaterials. However, although efforts have been made [7-9], reducing the serious Joule losses associated with the plasma resonances still remains challenging. Recently, all-dielectric metamaterials have become increasingly popular, which are composed of periodic or no periodic dielectric resonators made of high index and low loss dielectric materials. With the advantages like low losses, simple configuration and good compatibility with standard microfabrication, all-dielectric metamaterials have emerged as a promising and powerful platform for manipulating electromagnetic waves ranging from microwave to optics. Over the past few years, substantial achievements have been obtained in the field of optical all-dielectric metamaterials and numbers of all-dielectric metadevices including reflector [10-12],magnetic mirror [13], absorber [14,15], Huygens' metasuface [16-18] and etc. have been designed and demonstrated.

While optical all-dielectric metamaterials have become a hot research spot, there is a small quantity of literatures reported in THz region, which is a unique and important spectral region with a multitude of potential applications in imaging, wireless communication, security detection, chemical identification, sensing and so on. The limited reports on all-dielectric metamaterials in THz region are mainly because the developed fabrications have difficulty to seek a good balance between the precision and cost. To date, there are mainly three kinds of fabrication methods, including silicon deep etching method [19-23], laser etching method [24] and spray granulation [25]. The most widely used method is silicon deep etching method, which is of high-precision, but suffers from high-cost and complex fabrication process. As for direct laser patterning method, it features a fast approach to all-dielectric metamaterial, but suffers from high-cost and blurring boundaries. Unlike the methods mentioned above, the cost-efficient of spray granulation method allows for the preparation of high-throughput ceramic spherical particles (such as titanium dioxide). However, these particles are difficult to be assembled, and thereby their applications are seriously limited. As a result, more simpler and practical method is highly desired.

On the other hand, as for all-dielectric metamaterials, once the material and geometry are chosen, the corresponding optical responses are fixed, which would restrict their practical applications. Driven by the desire to realize the active tuning or modulating the optical properties of all-dielectric metamaterials, various approaches, including phase-change-media-based resonators [26-28], liquid crystals [29,30], flexible substrates [31], thermo-optic effect [32,33], doping [34] and etc. have been proposed in optical range. However, tunable all-dielectric metamaterials operating in THz range are rarely reported. Some attempts have been made including thermally tunable all-dielectric metamaterials based on strontium titanate rods and mechanically tunable all-dielectric metamaterials based on flexible substrates [24, 35].

Here, an effective method has been developed to prepare high-quality microspheres based all-dielectric metamaterial operating in terahertz (THz) range. Then, we propose a magnetically tunable all-dielectric metamaterial in THz range based on liquid crystal (LC). We demonstrate that the Mie resonances induced in the all-dielectric metamaterial can be tuned by external magnetic field and the tunability is highly sensitive to the directors of external magnetic field. A good tunability performance is obtained for the first electric resonance. We believe that the proposed method provides a promising way for tunable all-dielectric metamaterial, which would find considerable applications in THz range.

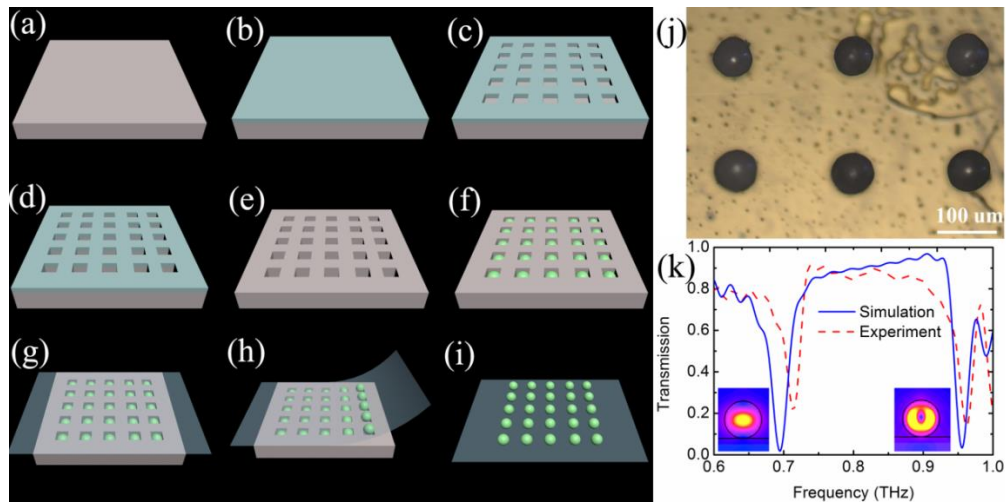

Fig. 1. (a)-(i) The fabrication process of all-dielectric metamaterial in THz range using template assistant method: (a) Standard cleaning process for silicon wafer. (b) Photoresist film deposition. (c) Lithographic technology process. (d) Deep silicon etching technology. (e) Removing photoresist film. (f) Filling of micro zirconia (yttria- stabilized)

spheres. (g) Pasting of PI tape. (h) Peeling off. (i) Fabricated all-dielectric metamaterial. (j) The microscope picture of fabricated all-dielectric metamaterial. (k) The simulated (blue solid line) and measured (red dash line) transmission in THz range. The inserts are the simulated magnetic field intensity distribution for the first and second resonances in x-z plane.

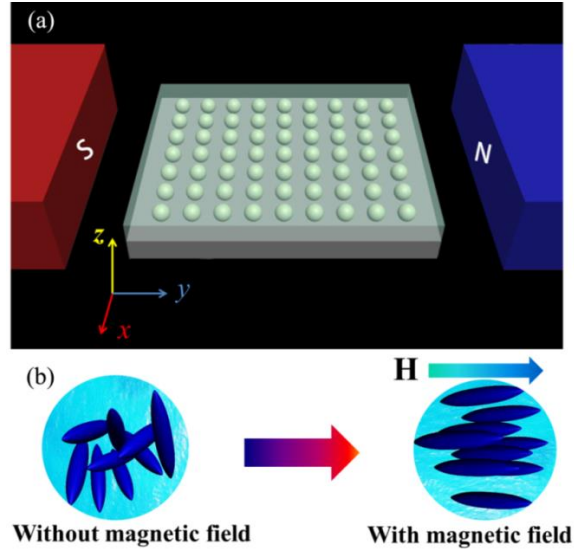

Fig. 2. (a) The model of LC- based active all-dielectric metamaterial under magnetic field. (b) The orientation of LC molecules with and without magnetic field.

In our study, the zirconia (yttria-stabilized) spherical particles are chosen as the building block of our metamaterial. By using inorganic sol–gel process, uniform zirconia (yttria-stabilized) spherical particles with a diameter from 50 to 150 um can be readily prepared according to our previous work [35]. To obtain high-quality all-dielectric metamaterial, a high-precision fabrication process, called template assistant method, is developed. The fabrication process flow can be seen in Fig. 1, which starts with the fabrication of silicon template. Firstly, the silicon substrate is cleaned by standard cleaning process (Fig. 1(a)). Then, 8 um thickness photoresist film is deposited on it (Fig. 1(b)) and developed with conventional lithographic technology (Fig. 1(c)). After that, by using inductively coupled plasma (ICP) etching machine, deep silicon etching technology is employed to obtain silicon template with hole arrays (Fig. 1(d)). The photoresist film is then removed (Fig. 1(e)) and the micro zirconia (yttria- stabilized) spheres fill the holes (Fig. 1(f)). To transfer the spheres to a flexible substrate, a polyimide (PI) tape is used to stick them (Fig. 1(g)). The PI tape is then peeled off (Fig. 1(h)) and a flexible all-dielectric metamaterial is obtained (Fig. 1(i)). It is worth mentioning that the geometrical parameters and packing arrangement of micro zirconia spheres can be adjusted by fabricating proper hole arrays with corresponding geometrical parameters and packing arrangement. In our study, the lattice constants in x direction and y direction are the same, $P_x = P_y = 210$ um. The diameter and height of the holes are 85 um and 100 um, respectively. The uniform micro zirconia (yttria- stabilized) spheres with a diameter of 80 um, little smaller than the one of holes, are chosen as dielectric resonators. The thickness of PI tape is 50 um. The micrograph of fabricated all-dielectric metamaterial is shown in Fig. 1(j), where the uniform micro zirconia (yttria- stabilized) spheres are highly ordered arrayed along the x direction and y direction.

The terahertz response of this metamaterial is measured by a photoconductive switch-based THz-TDS system [35]. In the measurement, the THz pulse passes through the sample under normal incidence and the polarization direction is along x direction. For comparison, simulation based on commercial soft, CST Microwave studio, is performed to obtain the corresponding transmission property. In the simulation, the permittivity and loss tangent of zirconia (yttria- stabilized) spheres are set as 35 and 0.01, respectively. The measured and simulated results are plotted in Fig. 1(k), where two pronounced transmission dips for the simulated spectrum can be observed at 0.694 THz and 0.956 THz, respectively. The measured first two dips appear at 0.715 THz and 0.964 THz, respectively. The simulated result and measured one show good agreement. To give a deeper insight into these two transmission dips, simulations are carried out to obtain the corresponding magnetic field intensity distribution. As depicted by the inserts in Fig. 1(k), the resonances can be determined as the first magnetic resonance and first electric resonance, respectively.

The magnetically tunable all-dielectric metamaterial is shown in Fig. 2(a), where the LC is sandwiched between two PI films (The top PI layer has not been shown here for simplification), and the all-dielectric resonators are immersed into LC. As is known, when the external magnetic field is large enough, the LC molecules are parallel to the magnetic field direction (Fig. 2(b)). Thereby, one can adjust the electromagnetic property of this metamaterial by changing the direction or intensity of external magnetic field. In the experiment, the thickness of the LC layer is 300 um, while the one for the top PI layer and bottom PI layer is 50 um. A commercial LC, 5CB, is utilized in our study. To produce a uniform magnetic field, parallel Nd–Fe–B sintered magnets are used. The magnetic field intensity between these two magnets is measured as 2000 G, a value much larger than the threshold field (about 100 G) required to reorientate the LC nematic molecules. It is worth mentioning that by adjusting the magnets, one can obtain a magnetic field with nearly arbitrary directions except for the one along the direction of propagation of light (along $z$ direction). To overcome this drawback, a solenoid is used to produce magnetic field with direction along the $z$ direction. We use a permittivity tensor ($\varepsilon_x$, $\varepsilon_y$, $\varepsilon_z$) to describe the electromagnetic properties of LC. Here, the permittivity elements for the wave perpendicular and parallel to the orientation of the LC molecules are $\varepsilon_\perp$ and $\varepsilon_{//}$, respectively. Assume that the LC molecules can be fully orientated by uniform magnetic field. When no external fields (magnetic field) is applied, the LC is an isotropic medium, which has a permittivity tensor expressed as ($\varepsilon_i$, $\varepsilon_i$, $\varepsilon_i$), where $\varepsilon_i = (2\varepsilon_\perp + \varepsilon_{//})/3$. When the magnetic field is set along $x$ direction, the permittivity tensor can be described as ($\varepsilon_{//}$, $\varepsilon_\perp$, $\varepsilon_\perp$). When the magnetic field is set along $y$ direction, the permittivity tensor can be described as ($\varepsilon_\perp$, $\varepsilon_{//}$, $\varepsilon_\perp$).

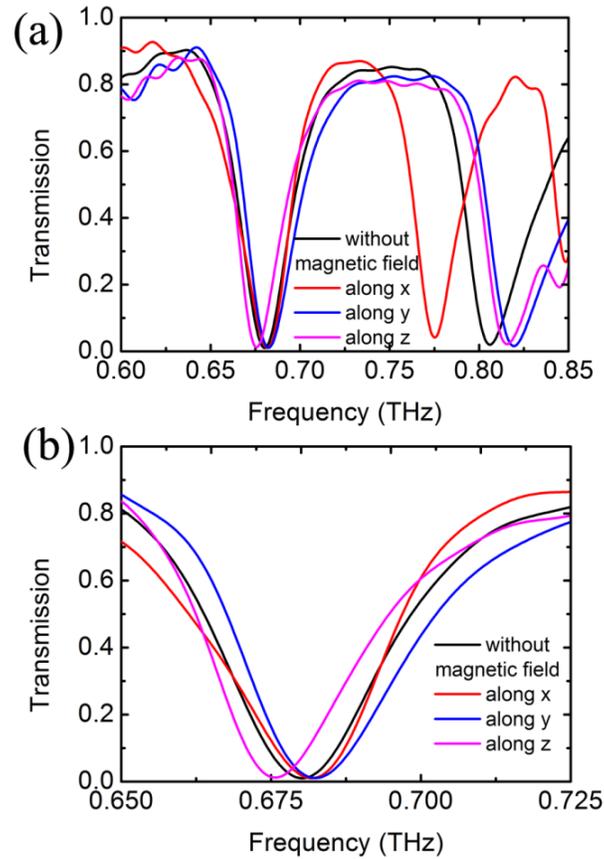

Fig. 3. (a) The simulated transmission of metamaterial with magnetic field and without magnetic field. (b) The enlarged drawing around the first resonance.

To predict the response of this metamaterial to the external magnetic field, simulations based on commercial software, CST Microwave Studio, are carried out. In the simulations, the THz beam with polarization direction along the $x$ direction passes through the sample under normal incidence. The complex permittivity of the PI is set as 3.0+0.001i. After reorientation, the LC has an ordinary index of 1.59 ($n_o$) and an extraordinary index of 1.74 ($n_e$). When the LC becomes isotropic, the refractive index can be described by $n_i = (2n_o + n_e)/3$ = 1.64. In our study, four cases are considered. Firstly, when no external magnetic field is applied, the LC is isotropic and its permittivity tensor can be described by ($\varepsilon_i$, $\varepsilon_i$, $\varepsilon_i$), where $\varepsilon_i = n_i^2$. The simulated transmission is plotted in Figs. 3(a) and 3(b), where one can find that the existing of LC makes the first magnetic and first electric resonances shift to lower frequencies. The first transmission dip associated to the first magnetic resonance locates at 0.672 THz, while the second one associated to the first electric resonance occurs at 0.804 THz. When the external magnetic field is applied along $x$ direction, the permittivity tensor is described by ($\varepsilon_{//}$, $\varepsilon_\perp$, $\varepsilon_\perp$). A remarkable redshift of 0.023 THz, from 0.804 THz to 0.781 THz, can be observed for the electric resonance. However, the magnetic resonance has a slight blueshift of 0.001 THz. When the external magnetic field is applied along $y$ direction, the permittivity tensor is given by ($\varepsilon_\perp$, $\varepsilon_{//}$, $\varepsilon_\perp$). The magnetic resonance shifts to a higher frequency of 0.674 THz (a blueshift of 0.002 THz), while the electric resonance moves to a higher frequency of 0.814 THz (a blueshift of 0.010 THz). Finally, when the external magnetic field is applied along $z$ direction, the permittivity tensor is described by ($\varepsilon_\perp$, $\varepsilon_\perp$, $\varepsilon_{//}$). The magnetic resonance shifts from 0.672 THz to 0.668 THz (a blueshift of 0.004 THz), while the electric resonance moves to 0.813 THz (a blueshift of 0.009 THz). As a result, as for magnetic resonance, a maximum tunability of 0.004 THz (from 0.672 THz to 0.668 THz) can be obtained when the direction of external magnetic field is set along $z$ direction. As for the electric resonance, a maximum tunability of 0.023 THz (from 0.804 THz to 0.781 THz) is achieved when the direction of external magnetic field is set along $x$ direction.

Using a THz-TDS system, we have measured the influence of external magnetic field on the property of this metadevice. In the experiment, the results are measured about 30 s after the external magnetic field is applied to make sure that the LC molecules have been fully orientated. As plotted in Fig. 4(a), the measured results show good agreement with the simulated ones, confirming the feasibility of the aforementioned prediction. When no magnetic field is applied, the measured first magnetic resonance and first electric resonance appear at 0.689 THz and 0.808 THz, respectively. When magnetic field is applied along $x$ direction, for the electric resonance, we experimentally obtain a maximum tunability of 0.021 THz (from 0.808 THz to 0.787 THz). A maximum absolute transmission modulation of 50.2 % (Fig. 4(b)) and a maximum modulation depth reaches 227% at 0.808 THz are obtained. As for magnetic resonance, we observe a maximum tunability of 0.003 THz (from 0.689 THz to 0.686 THz) when the magnetic field is applied along $z$ direction. The tunability in the simulation is slightly better than the one in the experiment, which can be attributed to two main reasons. The first one is the fabrication. The second one is the imperfect orientation of the LC molecules near the vicinity of resonators and PI films although strong magnetic field is applied.

At this point, we have investigated the magnetic tunability of LC-based all-dielectric metamaterial by changing the orientation of the LC molecules from isotropic state to an anisotropic one. In addition, strongly different tuning properties can be observed for the magnetic and electric resonances. As for practical applications, better tunability can be obtained by changing the orientation from an anisotropic state to another anisotropic one. This can be achieved by changing the direction of external magnetic field. For example, as for magnetic resonance, a maximum tunability of 0.006 THz (from 0.674 THz to 0.668 THz) can be observed when the direction of external magnetic field changes from $z$ direction to $y$ direction. We also calculate the absolute transmission modulation in such case, where a maximum absolute transmission modulation of 20% is achieved at 0.674 THz. As for the electric resonance, a maximum tunability of 0.033 THz (from 0.814 THz to 0.781 THz) is obtained when the direction of external magnetic field changes from $y$ direction to $x$ direction. Importantly, we observe a maximum absolute transmission modulation of 70 % at 0.781 THz. Experimentally, the maximum tunability of electric resonance is 0.036 THz (from 0.823 THz to 0.787 THz), while the one of magnetic resonance is 0.003 THz. A tuning figure of merit (FOM) is introduced to quantify the resonance switching, which is given by:

$$\text{FOM} = \frac{\text{tuning range (THz)}}{\text{FWHM (THz)}} \quad (1)$$

Here, FWHM is the full width at half maximum of the resonance. The larger FOM is, the better tunability is. Here, the FOM for the electric resonance reaches 0.9, suggesting the good magnetical tunability of this metadevice. Note that the tunability can be improved through optimization.

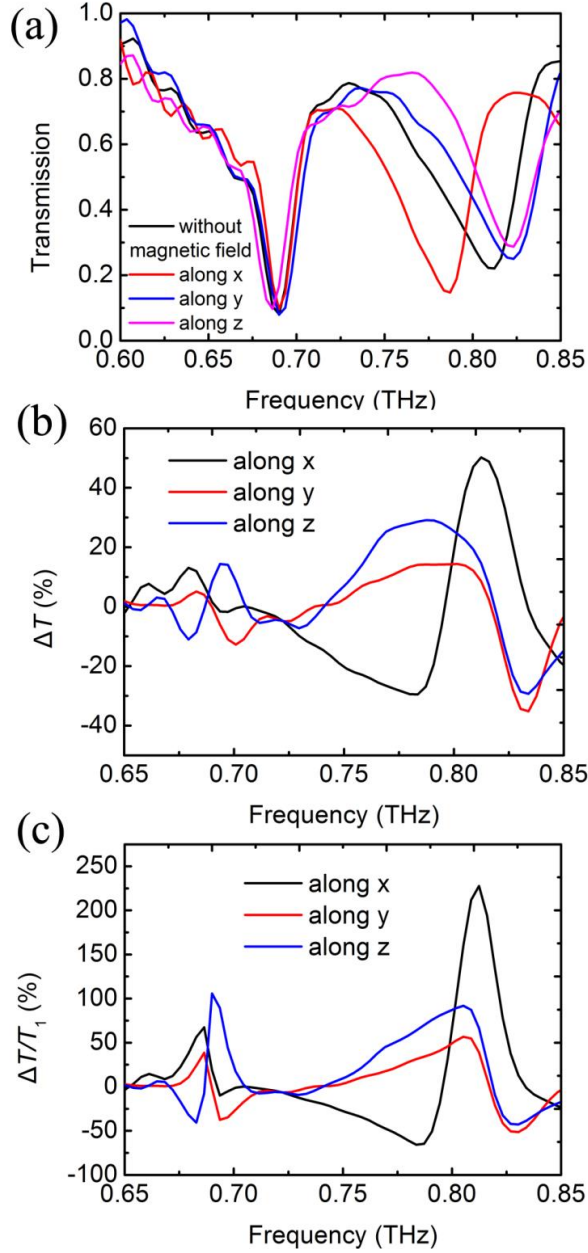

Fig. 4. (a) The measured transmission of metamaterial with magnetic field and without magnetic field. (b) The calculated absolute transmission modulation. (c) The calculated modulation depth.

In conclusion, a magnetically tunable all-dielectric metamaterials in THz range based on liquid crystal (LC) has been proposed and demonstrated. An effective method has been proposed to prepare high-quality all-dielectric metamaterial. The resonances can be effectively tuned by external magnetic field and the tunability is highly sensitive to the directors of magnetic field. A good tunability performance has been demonstrated for the first electric resonance. This work provides a promising way for tunable all-dielectric metamaterial, which would find considerable application in THz range.

**Funding**

This work is supported by the 111 project（NO.B17007）and Director Funds of Beijing Key Laboratory of Network System Architecture and Convergence. Fundamental Research Funds for the Central Universities (2018XKJC05), China Postdoctoral Research Foundation (Grant No. 2017M620693).